\documentclass[preprint,5p,twocolumn]{elsarticle}

\usepackage[utf8]{inputenc}
\usepackage{dcolumn}
\usepackage{amsmath}
\usepackage{amsfonts,amssymb,bm,tensor,braket}
\usepackage{xcolor}
\usepackage{autobreak}
\usepackage{graphicx}
\usepackage{threeparttable,array,booktabs}
\usepackage[referable]{threeparttablex}
\usepackage{makecell, multirow}
\usepackage{tabularx}
\usepackage{setspace}
\usepackage[varg]{txfonts}
\usepackage{caption}
\usepackage{hyperref}
\usepackage[nameinlink,noabbrev]{cleveref}
\journal{Science Bulletin, published as Sci. Bull. 66 (2021) 2254-2256, \url{https://doi.org/10.1016/j.scib.2021.07.030}}

\begin{document}

\begin{frontmatter}

%% Title, authors and addresses

%% use the tnoteref command within \title for footnotes;
%% use the tnotetext command for theassociated footnote;
%% use the fnref command within \author or \address for footnotes;
%% use the fntext command for theassociated footnote;
%% use the corref command within \author for corresponding author footnotes;
%% use the cortext command for theassociated footnote;
%% use the ead command for the email address,
%% and the form \ead[url] for the home page:
%% \title{Title\tnoteref{label1}}
%% \tnotetext[label1]{}
%% \author{Name\corref{cor1}\fnref{label2}}
%% \ead{email address}
%% \ead[url]{home page}
%% \fntext[label2]{}
%% \cortext[cor1]{}
%% \address{Address\fnref{label3}}
%% \fntext[label3]{}

\newcommand{\e}{{\mathrm{e}}}
\renewcommand{\i}{{\mathrm{i}}}
\renewcommand{\deg}{^\circ}

%% use optional labels to link authors explicitly to addresses:
%% \author[label1,label2]{}
%% \address[label1]{}
%% \address[label2]{}
\newcommand*{\PKU}{School of Physics and State Key Laboratory of Nuclear Physics and Technology, Peking University, Beijing 100871, China}
\newcommand*{\CHEP}{Center for High Energy Physics, Peking University, Beijing 100871, China}
\newcommand*{\CIC}{Collaborative Innovation Center of Quantum Matter, Beijing, China}

\title{LHAASO discovery of highest-energy photons towards new physics}

\author[a]{Chengyi Li}
\author[a,b,c]{Bo-Qiang Ma\corref{cor1}}

\address[a]{\PKU}
\address[b]{\CHEP}
\address[c]{\CIC}
%\address[d]{\CHPS}
\cortext[cor1]{Corresponding author \ead{mabq@pku.edu.cn}} %To publisher: please don't add additional contact information in publication.

\begin{abstract}
%Text of Abstract
Ultrahigh-energy photons up to 1.4 peta-electronvolts have been observed by new cosmic-ray telescope in China~\cite{Aharonian:2021pre} -- a hint that Lorentz invariance might break down at the Planck-scale level~\cite{Li:2021tcw}.
\end{abstract}

\begin{keyword}
%% keywords here, in the form: keyword \sep keyword

ultrahigh-energy gamma-ray, Lorentz invariance violation, quantum gravity, string theory model, space-time foam
%% PACS codes here, in the form: \PACS code \sep code

%% MSC codes here, in the form: \MSC code \sep code
%% or \MSC[2008] code \sep code (2000 is the default)

\end{keyword}

\end{frontmatter}

As one of the most sensitive gamma-ray detector arrays currently operating in the ultrahigh-energy~(UHE) band, the Large High Altitude Air Shower Observatory~(LHAASO) recently reported large numbers of gamma-ray photons with energies larger than $ 100$~TeV from twelve cosmic accelerators within the Milky Way~\cite{Aharonian:2021pre}, including the most high-energetic $\gamma$-ray detected at about $1.4$~PeV. The observation of such high-energy lights from space can be used not only to investigate the long-lasting puzzle of the origin of UHE cosmic rays and their acceleration mechanism in the extreme regions of the Universe, but can provide also the unique and crucial opportunity in testing fundamental concepts of physics, such as one basic symmetry underlying Einstein's relativity -- Lorentz invariance~\cite{Li:2021tcw}.

Set more than 4.4 km above sea level on the Haizi Mountain in Sichuan, China, the LHAASO is a new-generation mountain observatory~\cite{Cao:2019lnn} with unprecedented capability to spot UHE particles that rain down on Earth from cosmos. Instead of tracking cosmic rays that carry electric charges and hence are always bent by cosmological magnetic fields, LHAASO aims to detect high-energy $\gamma$-radiation photons, as they travel through the Universe in straighter lines and are relatively easier to be traced. The observatory consists of three interconnected arrays of detectors, the Kilometer Square Array~(KM2A), the Water Cherenkov Detector Array~(WCDA), and the Wide Field-of-view~(air) Cherenkov Telescope Array~(WFCTA), sprawling over a total area of about 1.36 km$^2$. Amongst these three components of the LHAASO instrument, KM2A was designed to measure ultrahigh-energy cosmic gamma-ray photons of energies above $10$~TeV. The construction of the observatory started in November 2017. It then began operations two years later, when LHAASO was not even half-complete. Specifically, the partly completed KM2A detection array has been operated stably to continuously collect the data since the end of 2019. However the LHAASO-KM2A instrument is still under construction and can be fully operational by July 2021.

\begin{figure}%[h]
\centering
\includegraphics[width = 8.6cm]{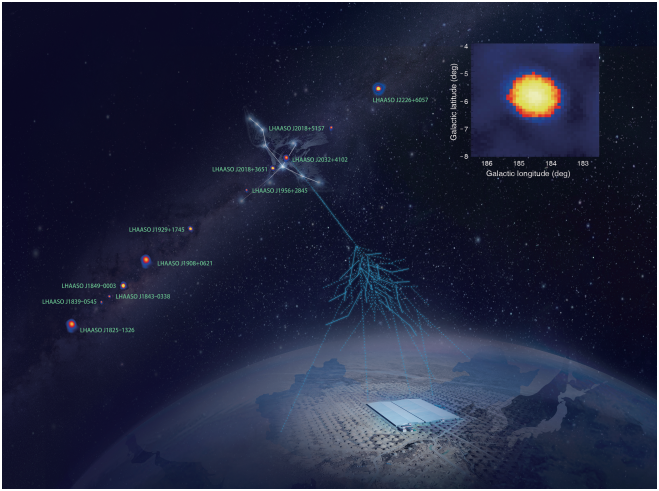}
\caption{(Color online) Ultrahigh-energy gamma-ray photons up to 1.4~PeV from twelve stable gamma-ray sources were detected by China's high-altitude observatory, LHAASO. Copyright $\copyright$ 2021, LHAASO Collaboration.\label{fig1}}
\end{figure}

Recently, the LHAASO collaboration~\cite{Aharonian:2021pre} announced the detection of more than 530 UHE photons coming from twelve astrophysical gamma-ray sources gathering around the Galactic disc~(\Cref{fig1}), using the KM2A half-array, with statistical significance of the photon signals from each source greater than $7\sigma$ over the surrounding background. These gamma-ray photons possess energies above 100~TeV up to the highest exceeding $1$~PeV, about a few dozen to a hundred times the particle energies that the Large Hadron Collider~(LHC) can produce, by noting that LHC reaches energies of $6.5$~TeV per particle in its second operational run. Amongst these LHAASO events, the most energetic photon with energy at $1.4$~PeV was found from the source LHAASO~J2032+4102, detected with a $10.5\sigma$ significance, and located in a very active star-forming region in the constellation of Cygnus. This record-breaking event turns out to be the highest energy photon ever observed by human being. These new observational results are of broad attention to the scientific society and are extremely encouraging for those studies aiming to search for the origins of UHE cosmic rays. Despite having several potential particle factories~(i.e., PeVatrons) within our Galaxy, including pulsar wind nebulae, remnants of supernova explosions, and star-forming regions~\cite{Aharonian:2021pre}, the cosmological objects responsible for these UHE gamma-rays have not yet been firmly localized and identified, leaving open the origin of these extreme cosmic accelerators.

The first discovery of cosmic gamma-rays at such high energies with LHAASO is of great significance. Since cosmic rays are high-energy particles from space, they may encode important information of the source objects. Precise measurements of these high-energy photons may provide clues to the astrophysical origins of cosmic rays and the processes that accelerate particles to such great speeds from the surrounding extreme regions and can give valuable hints on the evolution of celestial bodies and the early history of the Universe. On the other hand, such high-energetic photons in the PeV range may also enable us to test fundamental principles of physics and probe new physics that goes beyond the standard model~(SM) of particle physics and general relativity~(GR) of gravitation, such as ideas of violation of Lorentz invariance~(LV)~\cite{Li:2021tcw,Shao:2010wk}.

In Einstein's theory of special relativity, the speed of light and the laws of physics appear to be the same in all frames. The underlying symmetry, Lorentz invariance, is also the one that is mandated in the SM of particle physics. However, some extensions of the SM, especially those quantum gravity~(QG) scenarios that try to reconcile quantum mechanics with GR, motivate the breaking down of exact Lorentz symmetry in the proximity of the Planckian regime. Such a symmetry violation can lead to observable consequences that could serve as QG imprints at lower energies~\cite{Shao:2010wk}. Due to LV impacts, for instance, photons would no longer propagate through space at the constant light speed, but at a velocity that depends on the energy of that photon. In addition, photons with energies above a certain threshold would decay away. Such photon self-decay process is generally forbidden in the SM, since it breaks energy-momentum conservation. The ordinary reaction that two photons can interact and produce an electron-positron pair with an existing threshold could also possess an anomaly~\cite{Li:2021cdz} relative to special relativity if Lorentz invariance does not hold. These new-physics effects are so miniscule that it is hard to probe them from most terrestrial experiments, even at the LHC, whilst the high-energy astrophysical observation, such as those carried out by the LHAASO instrument, provides us an unprecedented laboratory to explore Lorentz symmetry violations.

Using the latest LHAASO data, we point out~\cite{Li:2021tcw} that the first detection of PeV gamma-rays may offer a very sensitive probe to test new physics of Lorentz violation. The detailed measurement of these UHE events by LHAASO allows us to constrain the photon decay motivated by superluminal type of Lorentz violation to a higher level~\cite{Li:2021tcw,Cao:2021pre,Chen:2021hen}, leading towards strong limits to certain Lorentz-violating frameworks. Quite remarkably, our result sets the tightest bound on the first-order Lorentz-violation energy scale ever reported, in excess of $\sim 2.7\times 10^{24}$~GeV, over about 200,000 times the Planck scale. This is deduced from the fact that LHAASO should not observe the 1.4~PeV photon that would undergo a rapid self-decay if superluminal Lorentz violation is allowed to take place. This result is very similar to that derived recently in Refs.~\cite{Cao:2021pre,Chen:2021hen}, using the same scenario.

Moreover, we indicate that a different type, i.e., subluminal type, of Lorentz invariance violation may be viable. Actually, a subluminal light speed variation at the scale of $\sim 3.6\times 10^{17}$~GeV has been proposed previously from analyses on gamma-ray burst~(GRB) data of multi-GeV photons~\cite{Xu:2016zxi,Xu:2016zsa,Xu:2018ien}. We further suggest that such type of subluminal deformation of Lorentz invariance is permitted to exist at the Planck-scale level in order to account for the latest LHAASO observation of PeV photons.  This is based upon the observation that subluminal Lorentz violation could drastically alter the conventional kinematics of photon-photon annihilation processes~\cite{Li:2021tcw}, as discussed also in Ref.~\cite{Li:2021cdz}. Thus these PeV gamma-rays would have chance to reach the Earth without attenuated by low-energy background lights diffusing in the sky~(e.g., cosmic background radiations) during their cosmological propagation. We propose~\cite{Li:2021tcw,Li:2021cdz} that any identification of such high-energy cosmic photons from extragalactic sources can be considered as support for Lorentz-violating properties of photons of the kind listed above.

On the theoretical side, we notice~\cite{Li:2021tcw} also a more fascinating implication for the LHAASO discovery, that is, certain models could account for all observed phenomenology on Lorentz violation within the photon sector, available to date, including the latest LHAASO result. In fact, it has been noted~\cite{Li:2021gah,Li:2021eza} that specific QG theories, especially the D-brane/string-inspired model for space-time foam~\cite{Ellis:2004ay,Ellis:2008gg,Li:2009tt} would be capable of explaining the light speed variation from GRB photons~\cite{Xu:2016zxi,Xu:2016zsa,Xu:2018ien}, in agreement with many other astrophysical data currently available. We further point out that this stringy QG framework may be also compatible with the data on the LHAASO UHE gamma-ray, since this type of models of (super)string theories naturally prevents the self decay for photons at high energies, but may predict an astrophysical signature of photon-photon annihilation threshold anomalies which, as explain above, exactly is what the new LHAASO experiment might indicate.

In the future, more observations with LHAASO could cast more stringent constraints to certain types~(e.g., superluminal type) of Lorentz invariance violation, but might also serve as supports for some Lorentz-violating QG schemes such as the string-inspired quantum space-time foam. These new incoming observations can thus offer hopefully valuable perspectives on the ultimate fate of the Planck-scale Lorentz symmetry.

\section*{Conflict of interest}
The authors declare that they have no conflict of interest.

\section*{Acknowledgments}
This work is supported by National Natural Science Foundation of China (Grant No.~12075003).

\section*{References}
%\noindent {\bf References}
%\vspace{3cm}

\end{document}